# A Reconfigurable Circuit Strategy and Its Application in Low-Power Rectifier for Ambient Energy Harvesting

Zhongqi He, *Member, IEEE*, Haoming He, *Graduate Student Member, IEEE*, Liping Yan, *Senior Member, IEEE*, and Changjun Liu, *Senior Member, IEEE*

*Abstract*—In ambient electromagnetic energy harvesting systems, the input power to the rectifier is low. To improve rectification efficiency, Schottky diodes, which are sensitive to low power, are commonly selected as rectifying devices to convert microwave power into dc power. However, low-power rectifying diodes typically have low reverse breakdown voltages, making them susceptible to reverse breakdown under high power conditions. This letter proposes a low-power rectifier with reconfigurable function. The rectifying diode is connected in parallel with the PIN diode. At low input power, the output dc voltage is low, and the PIN diode remains off, having no impact on the rectifier's operation. As the input power increases, the PIN diode turns on, causing change in circuit structure and impedance mismatch. This leads to increased reflected power, thereby preventing the rectifying diode from receiving excessive power. Additionally, the turn-on voltage of the PIN diode is lower than the reverse breakdown voltage of the rectifying diode, protecting it from reverse breakdown.

## I. INTRODUCTION

With the rapid development of wireless communication technologies, the amount of electromagnetic energy in the environment is increasing. At the same time, a large number of wireless sensors are being used in the Internet of Things (IoT). Currently, most wireless sensors are powered by batteries, and their ability to function normally is dependent on battery life. Regularly replacing batteries leads to increased costs and pollution issues. Additionally, some wireless sensors are located in areas that are inaccessible to humans, making battery replacement difficult. Therefore, harvesting electromagnetic energy from the environment to power wireless sensors could effectively address the power supply issue of these wireless sensors[1][2].

Rectifiers can convert microwave power into dc power [3][4][5], and are the core component of an electromagnetic energy harvesting system. Given that the energy density of electromagnetic waves in the environment is relatively low, the power input to the rectifier is also low. In order to achieve high conversion efficiency at low power levels, rectifying diodes are typically very sensitive to power. However, when close to a signal source, the received power increases rapidly, causing the rectifier's output voltage to rise. This may result in the rectifying diode being reverse-biased beyond its breakdown voltage, leading to the failure of the rectifier, as shown in Fig. 1.

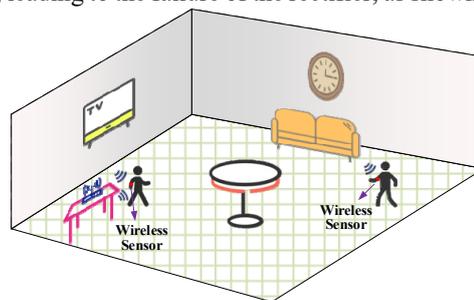

Fig. 1. Electromagnetic energy harvesting powers wireless sensors.

Some researchers have employed switches or additional rectifying diodes to increase the reverse breakdown voltage of the rectifying diodes. In [6], By using FET switches to reconfigure the rectifier, the reverse breakdown voltage of the diode is increased in the rectifier. Ref. [7] proposed a synergistic structure that introduces a high-power diode in front of the low-power diode to prevent it from breakdown. A recently proposed power-adaptive distribution structure enables the low-power rectifying diode to operate under low-power conditions, while the high-power rectifying diode functions under high-power conditions[8].

Currently, there is limited research on the protection of low-power rectifiers. This letter proposes a novel reconfigurable mechanism for low-power rectifying diodes by introducing a PIN diode as a dc switch. This mechanism adaptively adjusts the switch based on the output voltage. When the PIN diode is turned on, the circuit structure is reconfigured, causing impedance mismatch. This prevents high input power from reaching the rectifying diode, protecting it from damage.

## II. PRINCIPLE OF THE RECONFIGURABLE STRATEGY

In the protection of low-power nonlinear components, the

Manuscript received ——, ——.
This work was supported in part by the Natural Science Foundation of China (NSFC) under Grant 62401382 and U22A2015, the National Funded Postdoctoral Researcher Program of China under Grant GZC20231766, and the Sichuan Science and Technology Program under Grant 2024YFHZ0282 and 2025ZNSFSC1441. (Corresponding author: *Changjun Liu*.)
The authors are with the School of Electronics and Information Engineering, Sichuan University, Chengdu 610064, China (e-mail: cjliu@ieee.org).

conventional and straightforward approach is to introduce a limiter. As shown in Fig. 2(a), when the input microwave power exceeds a certain threshold, the output microwave power remains constant, thereby protecting the subsequent circuits.

Different from the traditional design method, this letter proposes a novel circuit structure that utilizes an adaptive structure regulation mechanism to control the power entering the rectifying diode and the output dc voltage, as shown in Fig. 2 (b). The switch state of the voltage-controlled component depends on the voltage across it. This component is connected in parallel with the rectifying diode. When the input power is low, the output dc voltage remains low, keeping the voltage-controlled component in the off state and leaving the rectifier's operation unaffected. As the input power gradually increases and the output dc voltage rises, the voltage-controlled component turns on, restructuring the circuit. This reconfiguration causes impedance mismatch and increased reflection, preventing excessive power from reaching the rectifying diode.

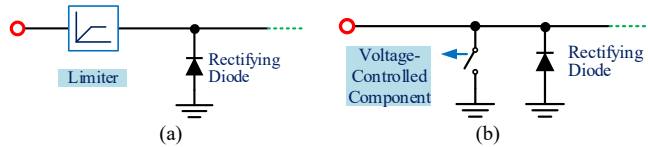

Fig. 2. Methods for protecting rectifiers. (a)Traditional low-power rectifier with a limiter, and (b)the proposed rectifier with a voltage-controlled component.

### III. DESIGN OF THE PROPOSED RECTIFIER

*A. DC Voltage at the Maximum Conversion Efficiency*

When the input microwave power is low, the rectifying efficiency is low due to the influence of the diode's turn-on voltage. As the power gradually increases, the conversion efficiency improves and eventually reaches its peak value, achieving the optimal dc voltage. However, as the power continues to rise, the efficiency of the rectifier rapidly decreases due to the impact of the reverse breakdown voltage, which may also lead to damage.

The goal of the reconfigurable strategy proposed in this letter is to maintain the maximum conversion efficiency of the rectifier; meanwhile, as the power continues to increase, prevent the rectifier from outputting excessively high voltage, thereby protecting the rectifying diode from reverse breakdown. Therefore, it is first necessary to study the dc voltage of the rectifier at its maximum conversion efficiency.

Fig. 3 is the nonlinear model of the rectifying diode [9], including the series resistance $R_s$, the junction resistance $R_j$, and the junction capacitance $C_j$.

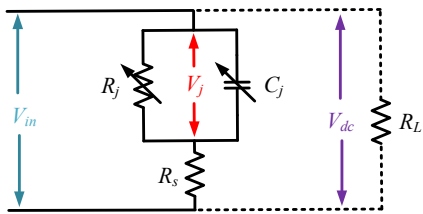

Fig. 3. Nonlinear model of the rectifying diode.

The input voltage $V_{in}$ can be written as:
$$V_{in} = -V_{dc} + V_{mw}\cos(\omega t) \quad (1)$$
where $V_{dc}$ is the dc voltage on the load, and $V_{mw}$ is the input microwave voltage.

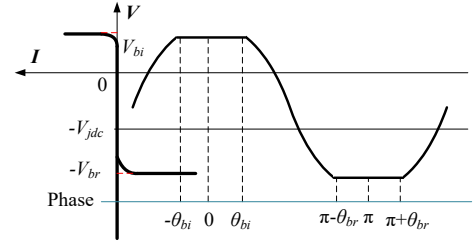

Fig. 4. I-V characteristic curve of the rectifying diode and the time-domain waveform of the junction voltage $V_j$.

According to Fig. 4, the junction voltage $V_j$ can be expressed as:
$$V_j = \begin{cases} V_{bi}, & V_j > V_{bi} \\ -V_{jdc} + V_{jmw}\cos(\omega t - \varphi_0), & V_{bi} > V_j > -V_{br} \\ -V_{br}, & -V_{br} > V_j \end{cases} \quad (2)$$

where $V_{bi}$ and $V_{br}$ are the diode's forward conduction voltage and reverse breakdown voltage, respectively, and $V_{jdc}$ and $V_{jmw}$ represent the dc component and the fundamental amplitude of the junction voltage when the diode operates in the reverse-biased state. In Fig. 4, $\theta_{bi}$ and $\theta_{br}$ are the forward conduction angle and reverse breakdown angle, respectively. Due to the presence of the junction resistance $R_s$, there exists a phase difference $\varphi_0$ between the input voltage $V_{in}$ and the diode junction voltage $V_j$.

When the diode junction voltage exceeds the forward conduction voltage, the diode operates in the conduction state, satisfying the following relationship:
$$V_{jmw} \times \cos\theta_{bi} - V_{jdc} = V_{bi} \quad (3)$$

When the diode junction voltage exceeds the reverse breakdown voltage, the diode operates in the breakdown state, satisfying the following relationship:
$$V_{jmw} \times \cos\theta_{br} + V_{jdc} = V_{br} \quad (4)$$

When the diode operates in the off state, the junction resistance $R_j$ is ideally infinite. The diode junction current $I_{off,Rs}$ flowing through $R_s$ entirely flows into the junction capacitance $C_j$, and is expressed as:
$$I_{off,Rs} = \frac{d(C_j V_j)}{dt} = \frac{V_{in} - V_j}{R_s} \quad (5)$$

Substituting equations (1) and (2) into the above equation (5), we obtain:
$$I_{off,Rs} = -\omega C_j V_{jmw}\sin\theta = (V_{jdc} - V_{dc}) + (V_{mw}\cos\varphi_0 - V_{jmw})\cos\theta \quad (6)$$
$$- V_{mw}\sin\varphi_0 \sin\theta$$

where $\theta = \omega t - \varphi_0$.

This equation has a solution and can be easily seen to satisfy the following relationship:
$$V_{jdc} = V_{dc} \quad (7)$$
$$V_{jmw} = V_{mw} \quad (8)$$

It should be noted that, since the microwave rectifying diode used typically has a small series resistance, the resulting diode voltage drop is also small [9]. Here, the phase difference $\varphi_0$ caused by it is approximated to be zero, i.e., $\varphi_0 \approx 0$.

Substituting (7) and (8) into (3) and (4), the following critical state relationship can be obtained:

$$\frac{V_{dc}+V_{bi}}{\cos\theta_{bi}} = \frac{V_{br}-V_{dc}}{\cos\theta_{br}} \quad (9)$$

Under ideal conditions, the power loss during microwave-to-dc conversion is primarily attributed to the power dissipation on the diode.

When the diode is in the forward conduction state, the junction resistance $R_j$ is approximately zero, and the conduction current is relatively high [9]. This conduction current flowing through the series resistance $R_s$ and the junction resistance $R_j$ results in power losses denoted as $L_{bi,Rs}$ and $L_{bi,Rj}$, respectively.

When the diode is in the reverse cutoff state, the junction resistance $R_j$ becomes infinite, and the leakage current flowing through $R_s$ results in power loss $L_{off,Rs}$.

When the diode is in the breakdown state, the breakdown current is also relatively high, leading to power losses $L_{br,Rs}$ and $L_{br,Rj}$ due to the current passing through $R_s$ and $R_j$. Table I summarizes the five types of losses in a rectifying diode.

TABLE I
POWER LOSSES ON THE RECTIFYING DIODE

|  | Conduction State | Cutoff State | Breakdown State |
|---|---|---|---|
| $R_s$ | $L_{bi,Rs}$ | $L_{off,Rs}$ | $L_{br,Rs}$ |
| $R_j$ | $L_{bi,Rj}$ | 0 | $L_{br,Rj}$ |

The forward conduction angle $\theta_{bi}$ must satisfy $0 < \theta_{bi} < \pi$, while the reverse breakdown angle $\theta_{br}$ must satisfy $0 \leq \theta_{br} < \pi$. The smaller the forward conduction angle $\theta_{bi}$ and the reverse breakdown angle $\theta_{br}$, the lower the power loss. When the rectifier reaches its maximum conversion efficiency, it implies that $\theta_{br} = 0$ and $\theta_{bi} \to 0$. This condition represents the critical state, which can be expressed as:

$$\cos\theta_{br}\big|_{\theta_{br}=0} = 1 \quad (10)$$

$$\cos\theta_{bi}\big|_{\theta_{bi}\to 0} \approx 1 \quad (11)$$

Substituting the above critical state conditions (maximum efficiency) into equation (9), it can be determined that when the rectifier operates at its highest efficiency, the dc output voltage $V_{dc}$ satisfies:

$$V_{dc} = \frac{V_{br}-V_{bi}}{2} \quad (12)$$

It can thus be concluded that when the microwave rectifier operates at its optimal state (maximum efficiency), the dc output voltage is approximately equal to half the difference between the absolute values of its breakdown voltage and forward conduction voltage.

This conclusion will guide the rectifier design proposed in this letter, and we will protect the rectifier without affecting the maximum conversion efficiency.

*B. Rectifier Design*

Fig. 5 shows the schematic diagram of the designed low-power rectifier. In series with the rectifying diode is an eighth-wavelength short-ended transmission line, whose input impedance is given by:

$$Z_{TL} = jZ_0 \tan\left(\frac{\pi}{4}\frac{\omega}{\omega_0}\right) = \begin{cases} 0, & \omega = 0 \\ jZ_0, & \omega = \omega_0 \\ \infty, & \omega = 2\omega_0 \end{cases} \quad (13)$$

where $Z_0$ is the characteristic impedance of the eighth-wavelength short-ended transmission line.

The eighth-wavelength short-ended transmission line exhibits an infinite impedance at the 2nd harmonic, effectively suppressing the 2nd harmonic and improving the rectification efficiency.

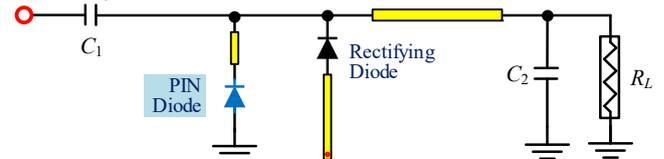

Fig. 5. Schematic of the proposed rectifier.

The rectifying diode is a Schottky diode SMS 7630, from Skyworks Solutions, Inc.. According to the datasheet of the SMS 7630, its forward turn-on voltage and reverse breakdown voltage are $V_{bi} = 0.34$ V and $V_{br} = 2$ V, respectively. As derived from Equation (12), when the diode SMS 7630 achieves its optimal conversion efficiency, the output dc voltage is approximately 0.8 V. The voltage-controlled switch in the proposed rectifier is two PIN diodes SMP 1345 (the conduction voltage of a single PIN diode is 0.89V, which will affect the maximum efficiency of the rectifier), from Skyworks. They are widely used in RF circuits.

According to the input power, the rectifier can be divided into two operating modes.

**Mode 1:** When the input power is low, the dc voltage output by the rectifying diode is low, and the PIN diode remains off. At this time, the microstrip line in series with the PIN diode is in an open-circuit state. The rectifier is in an impedance-matched state and operates efficiently.

**Mode 2:** When the input power is high, the rectifying diode outputs a higher voltage, causing the PIN diode to turn on. At this point, the microstrip line in series with the PIN diode transitions from an open-circuit state to a short-circuit state. This results in impedance mismatch in the rectifier, increasing reflection and reducing the power entering the rectifying diode, thereby protecting the rectifying diode from receiving excessive power. Additionally, the PIN diode limits the maximum output dc voltage—specifically, the output voltage cannot exceed the PIN diode's turn-on voltage. This effectively protects the rectifying diode from reverse breakdown under high-power conditions.

IV. SIMULATION, AND MEASUREMENT

To verify the proposed reconfigurable strategy for the low-power rectifier, we fabricated the proposed rectifier operating at 2.45 GHz. Its layout and the photograph are shown in Fig. 6(a). The rectifier is optimized around 0 dBm, and the measured $|S_{11}|$ is shown in Fig. 6(b). when the power is lower than 2 dBm, the reflection coefficients of the protected and unprotected rectifiers match well. However, as the input power increases, the PIN diodes in the protected rectifier turn on, restructuring the circuit and causing impedance mismatch. This leads to increased reflection (reflected power increased fivefold @ 8dBm), preventing excessive power from entering the rectifying diode and protecting it from breakdown.

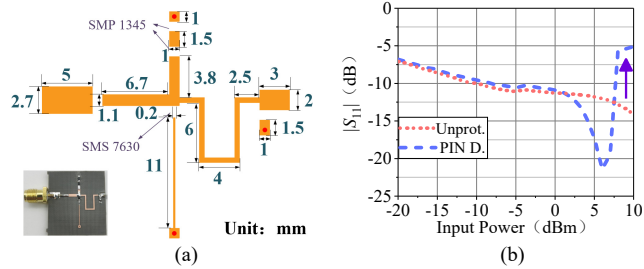

Fig. 6. (a) Layout and photograph of the proposed rectifier. (b) Measured $|S_{11}|$.

The traditional circuit protection method using limiters is employed for comparison. We selected two limiters from Mini-Circuits: the ZFLM-252-1WL-S+ (limiter 1) [11] and the VLM-33W-2W-S+ (limiter 2) [12]. These limiters can restrict the microwave power at 2.45 GHz to within -2 dBm and 10 dBm, respectively. Their insertion losses are approximately 4 dB and 0.6 dB around the input power level of 0 dBm, respectively.

The measurement results of the rectifier without protection, with two PIN diodes, the limiters 1 and 2, respectively, are shown in Fig. 7. When the output dc voltage is around 0.8 V, the unprotected rectifier achieves its peak efficiency of 69.8% at 2 dBm, which is consistent with the analysis presented in Section III. As the input power continues to increase, the dc voltage will reach the reverse breakdown voltage, leading to circuit damage.

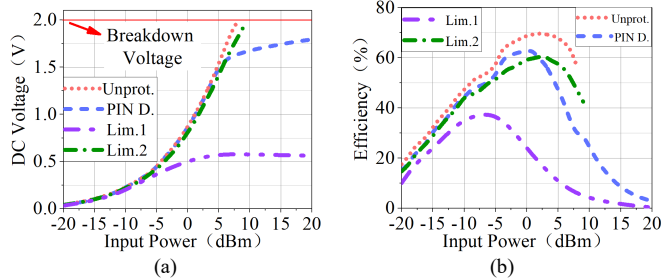

Fig. 7. Measured (a) dc voltage and (b) efficiency of the rectifier without protection, with two PIN diodes, the limiters 1 and 2, respectively.

Limiter 1 can successfully keep the voltage within a safe range; however, its high insertion loss results in very low efficiency. Limiter 2 fails to keep the voltage within a safe range and also introduces insertion loss.

While, the proposed reconfigurable strategy hardly reduces the efficiency of the rectifier at low power levels. Even when the input power increases (up to 0 dBm), the rectification efficiency only decreases by less than 6%. More importantly, even if the input power reaches 20 dBm (a power that is almost impossible to receive), the rectifier can still limit the voltage within a safe range.

In terms of circuit cost, the PIN diode is much cheaper than a limiter. Therefore, this design offers advantages over traditional designs in both performance and cost.

TABLE II
COMPARISON WITH RECTIFIERS WITH LOW-POWER DIODE PROTECTION

| Ref. | Year | Freq. (GHz) | Eff. @0dBm | Max. Eff. | Protection Method | Voltage Limit | Size (mm$^2$) |
|---|---|---|---|---|---|---|---|
| [7] | 2019 | 2.4 | 53% | 60.1% @10dBm | Cooperative Structure | No | 55 ×38 |
| [5] | 2024 | 2.4 | 54% | 71.2% @12dBm | Two-Stage Rectifier | No | 40 ×35 |
| [8] | 2024 | 2.43 | 53.4% | 69% @14dBm | Power Division | No | 39 ×35 |
| **This work** | **2025** | **2.45** | **62.8%** | **63% @1dBm** | **Reconfigurable Circuit** | **Yes** | **30 ×21** |

As shown in Table II, this design demonstrates advantages in terms of efficiency in low power, voltage limit, the number of dc loads, and circuit size.

## V. CONCLUSION

A low-power rectifier protection strategy for electromagnetic energy harvesting applications was designed and validated in this letter. By incorporating a PIN diode to dynamically adjust the structure of the rectifier, the rectifier is effectively protected from reverse breakdown with almost no impact on the rectification efficiency. Moreover, the proposed protection strategy can also be applied to other circuits that are at risk of reverse breakdown and high power.